\documentclass[9pt,times,mathptm,psfig,twocolumn]{IEEEtran}

\usepackage{algorithmicx}
\usepackage{algorithm}
\usepackage{algpseudocode}
\usepackage{multirow}
\usepackage{subcaption}
\usepackage{setspace}
\usepackage[utf8]{inputenc}
\usepackage{amsmath}
\usepackage{amsfonts}
\usepackage{amssymb}
\usepackage{graphicx}
\begin{document}


\title{Efficient Parallel Verification of Galois Field Multipliers}

\author{Cunxi Yu, ~Maciej Ciesielski \\
ECE Department,~University of Massachusetts, Amherst, USA \\
    ycunxi@umass.edu, ~ciesiel@ecs.umass.edu}
\maketitle

{\it\bf Abstract -} 
Galois field (GF) arithmetic is used to implement critical arithmetic components in communication and security-related hardware, and verification of such components is of prime importance. Current techniques for formally verifying such components are based on computer algebra methods that proved successful in verification of integer arithmetic circuits. However, these methods are sequential in nature and do not offer any parallelism. This paper presents an algebraic functional verification technique of gate-level $GF(2^m)$ multipliers, in which verification is performed in bit-parallel fashion. The method is based on extracting a unique polynomial in Galois field of each output bit independently. We demonstrate that this method is able to verify an \textit{n}-bit GF multiplier in \textit{n} threads. Experiments performed on pre- and post-synthesized \textit{Mastrovito} and \textit{Montgomery} multipliers show high efficiency up to 571 bits.
 \\


\section{Introduction}

Galois field (GF) arithmetic is used to implement critical arithmetic components in communication and security-related hardware. It has been extensively applied in many digital signal processing and security applications, such as Elliptic Curve Cryptography (ECC), Advanced Encryption Standard (AES), and others. Multiplication is one of the most heavily used Galois field computations and it is a high complexity operation. Specifically, in cryptography systems, the size of Galois field circuits can be very large. Therefore, developing general formal analysis technique of Galois field arithmetic HW/SW implementations becomes critical.  Contemporary formal techniques, such as \textit{Binary Decision Diagrams} (BDDs), \textit{Boolean Satisfiability} (SAT), \textit{Satisfiability Modulo Theories} (SMT), etc., are not directly applicable to either the verification or reverse engineering of Galois field arithmetic. The limitations of these techniques when applied to Galois field arithmetic have been addressed in \cite{kalla:tcad13}. 

The most successful techniques for verifying arithmetic circuits use computer algebra techniques with polynomial representations \cite{kalla:tcad13}\cite{STABLE:date11}\cite{sayedformal:date-2016}\cite{ciesielski2015verification}. The verification problem is typically formulated as proving that the implementation satisfies the specification. This is accomplished by performing a series of divisions of the specification polynomial $F$ by the implementation polynomials $B=\{f_1, \dots, f_s\}$, representing components that implement the circuit. The technique based on \textit{Gr{\" o}bner Basis} demonstrated that this approach can efficiently reduce the complexity of the verification problem to \textit{membership testing} of the specification polynomial in the ideals \cite{kalla:tcad13}\cite{sayedformal:date-2016}. This technique has been applied successfully to large Galois Field arithmetic circuits \cite{kalla:tcad13}. Symbolic computer algebra methods have been used to derive word-level operation for GF circuits and integer arithmetic circuits to improve the verification performance \cite{kalla:dac2014}\cite{yu:2016-abstraction}. A different approach to arithmetic verification of synthesized gate-level circuits has been proposed in \cite{ciesielski2015verification}. This method uses algebraic rewriting of the polynomials at the primary outputs to extract specification as a polynomial at the primary inputs.

However, a common limitation to all these works is that they are not applicable to parallel verification. This is because the verification problem based on computer algebra technique expresses the \textit{specification} as polynomial in all output bits. In this approach, the polynomial division can be done only in a single thread. 
In principle, multiple specifications (called \textit{output signature} in \cite{ciesielski2015verification}) can be generated by splitting the output signature. However, we examined this method and found the performance to be really poor. The reason is that the technique of \cite{ciesielski2015verification} needs to rewrite the entire output signature in all the output bits to benefit from large monomial cancellations during rewriting. In other works \cite{kalla:tcad13}\cite{kalla:dac2014}, the verification problem of post-synthesized Galois field multipliers have not been addressed. 

In this work, we extend the verification technique of \cite{ciesielski2015verification} to verification of Galois field multipliers, while applying bit-level parallelism. Specifically:
\begin{itemize}
\item We propose an algorithm for Galois field arithmetic verification, which significantly reduces the internal expression size during algebraic rewriting.
\item We evaluate our approach using benchmarks used in \cite{kalla:tcad13}\cite{kalla:dac2014}, including \textit{Mastrovito} and \textit{Montgomery} multipliers, up to 571 bits. The results show that efficiency of our approach surpasses that of \cite{kalla:tcad13} and \cite{kalla:dac2014}.
\item We demonstrate that for the verification problem for an $n$-bit Galois field multiplier can be accomplished ideally in $n$ parallel threads. In this work, we set the number of threads to 5, 10, 20, and 30. We also analyze the efficiency of our parallel approach by studying the tradeoff between CPU runtime and memory usage.
\item We address the verification of {\it synthesized} Galois field multipliers, while previous work dealt only with the verification of structural representation (arithmetic netlists) prior to synthesis.
\end{itemize}

\section{Background}

Different variants of canonical, graph-based representations have been proposed for arithmetic circuit verification, including Binary Decision Diagrams (BDDs) \cite{bryant:1986-bdd}, Binary Moment Diagrams (BMDs) \cite{bmd95}, Taylor Expansion Diagrams (TED) \cite{ted:tcomp06}, and other hybrid diagrams. 
While the canonical diagrams have been used extensively in logic synthesis, high-level synthesis and verification, their application to verify large arithmetic circuits remains limited by the prohibitively high memory requirement for complex arithmetic circuits \cite{ciesielski2015verification}\cite{kalla:tcad13}.  Alternatively, arithmetic verification problems can be modeled and solved using Boolean satisfiability (SAT) or satisfiability modulo theories (SMT). However, it has been demonstrated that these techniques cannot efficiently solve the verification problem of large arithmetic circuits \cite{kalla:tcad13} \cite{cunxi:2016-tcad-verification}. Another class of solvers include Theorem Provers, deductive systems for proving that an implementation satisfies the specification, using mathematical reasoning. However, this technique requires manual guidance, which makes it difficult to be applied automatically.

\subsection{Computer Algebra Approaches}

The most advanced techniques that have potential to solve the arithmetic verification problems are those based on symbolic Computer Algebra. 
These methods model the arithmetic circuit specification and its hardware implementation as polynomials  \cite{kalla:tcad13}\cite{STABLE:date11}\cite{ciesielski2015verification}\cite{kalla:dac2014}\cite{wienand:cav08}. 
The verification goal is to prove that implementation satisfies the specification by performing a series of divisions of the specification polynomial $F$ by the implementation polynomials $B=\{f_1, \dots, f_s\}$, representing components that implement the circuit. 
%
The polynomials $f_1,...,f_s$ are called the bases, or {\it generators}, of the ideal $J$. 
Given a set $f_1,...,f_s$ of generators of $J$, 
a set of all simultaneous solutions to a system of equations $f_1(x_1,...,x_n)$=0; ...,$f_s(x_1,...,x_n)$=0 is called a {\it variety} $V(J)$. 
Verification problem is then formulated as testing if the specification $F$ vanishes on $V(J)$ 
In some cases, the test can be simplified to checking if $F \in J$, which is known in computer algebra as {\it ideal membership} testing \cite{kalla:tcad13}. 

There are two basic techniques to reduce polynomial $F$ modulo $B$. A standard procedure to test if $F \in J$ is to divide polynomial $F$ by the elements of $B$: $f_1,...,f_s$, one by one. The goal is to cancel, at each iteration, the leading term of $F$ using one of the leading terms of $f_1,...,f_s$.
If the remainder of the division is $r=0$, then $F$ vanishes on $V(J)$, proving that the implementation satisfies the specification. However, if $ r \ne 0 $, such a conclusion cannot be made: 
$B$ may not be sufficient to reduce $F$ to 0, and yet the circuit may be correct. To check if $F$ is reducible to zero, a {\it canonical} set of generators, $G=\{g_1,...,g_t\}$, called {\it Gr{\" o}bner basis} is needed. This technique has been successfully applied to Galois field arithmetic \cite{kalla:tcad13} and integer arithmetic circuits \cite{sayedformal:date-2016}. A different approach has been proposed in \cite{ciesielski2015verification}\cite{yu-iscas15}\cite{yu:2016-abstraction}\cite{samaneh:2015-debug}\cite{yu-isvlsi-16a}, where a gate-level network is described by a system of equations and proved by \textit{backward rewriting}. Starting with the known output signature (polynomial) in primary output variables, it rewrites the signature from the primary outputs to primary inputs, to extract an arithmetic function (specification). The specific verification work of Galois field arithmetic has been presented in \cite{kalla:tcad13} \cite{kalla:dac2014}. These works provide significant improvement compared to other techniques, since their formulations relies on certain simplifying properties in Galois field during polynomial reductions. Specifically, the problem reduces to the ideal membership testing over a larger ideal that includes $J_0 = \langle x^2-x \rangle $ in ${\mathbb{F}}_2$. In this paper, we provide a comparison between this technique and our approach.


\subsection{Galois Field Multiplication}

Galois field (GF) is a number system with a finite number of elements and two main arithmetic operations, addition and multiplication; other operations can be derived from those two \cite{paar2009understanding}. Galois field with $p$ elements is denoted as $GF(p)$. The most widely-used finite fields are \textit{Prime Fields} and \textit{Extension Fields}, and particularly {\it binary extension fields}. Prime field, denoted $GF(p)$, is a finite field consisting of finite number of integers \{$1,2, ....,p-1$\}, where $p$ is a prime number, with additions and multiplication performed \textit{modulo p}. 
Binary extension field, denoted $GF(2^m)$ (or $\mathbb{F}_{2^m}$), is a finite field with $2^m$ elements; unlike in prime fields, however, the operations in extension fields is not computed \textit{modulo $2^{m}$}. Instead, in one possible representation (called polynomial basis), 
each element  of $GF(2^m)$ is a {\it polynomial ring} with $m$ terms with the coefficients in $GF(2)$.  Addition of field elements is the usual addition of polynomials, with coefficient arithmetic performed modulo 2.  For example, a 2-bit vector $A$=\{$a_{0},~a_{1}$\} in $GF{(2^{2})}$, is $A(x)$=$a_{0}$+$a_{1}x$, where $a_{i} \in$ $GF(2)$=\{0,1\}.
Multiplication of field elements is performed modulo {\it irreducible polynomial} $P(x)$ of degree $m$ and coefficients in $GF(2)$. For example, $P$=$x^{2}+x+1$ is an irreducible polynomial in $GF{(2^{2})}$. The irreducible polynomial $P(x)$ is analog to the prime number $p$ in prime fields $GF(p)$. 
%
Extension fields are used in many cryptography applications, such as AES and ECC. In this work, we focus on the verification problem of $GF{(2^{m})}$ multipliers.

\begin{figure}[!htb]
    \begin{minipage}{.6\linewidth}
\centering
 \scriptsize
\begin{tabular}{cccc}
   &      & $a_1$   & $a_0$   \\
   &      & $b_1$   & $b_0$   \\ \hline
   &      & $a_1b_0$ & $a_0b_0$ \\
   & $a_1b_1$ & $a_0b_1$ &      \\ \hline
$r_3$ & $r_2$   & $r_1$   & $r_0$  
\end{tabular}
    \end{minipage}%
 \begin{minipage}{.4\linewidth}
\begin{tabular}{ccc}
 & $a_1$ & $a_0$ \\
 & $b_1$ & $b_0$ \\ \hline
 & $a_1b_0$ & $a_0b_0$ \\
$a_1b_1$& $a_0b_1$ &  \\ \hline
\multicolumn{1}{l}{$~s_2$} & \multicolumn{1}{l}{$~s_1$} & \multicolumn{1}{l}{$~s_0$} \\
\multicolumn{1}{l}{} & \multicolumn{1}{l}{$~s_2$} & \multicolumn{1}{l}{$~s_2$} \\ \hline
\multicolumn{1}{l}{} & \multicolumn{1}{l}{$~z_1$} & \multicolumn{1}{l}{$~z_0$}
\end{tabular}
\end{minipage} 
\hspace*{25mm } a) \hfill b) \hspace*{15mm}
\caption{2-bit multiplication: a) standard integer multiplication with 4-bit result; b) multiplication in $GF(2^2)$ with $A(x)$ = $a_{0}$+$a_{1}x$, $B(x)$ = $b_{0}$+$b_{1}x$ and result  $Z(x)$ = $z_{0}$+$z_{1}x$ $\equiv$ $A(x)$$\cdot$$B(x)$ mod $P(x)$; irreducible polynomial $P(x)=x^2+x+1$.} 
\label{fig:2-bit}
\end{figure}

An example of multiplication in $GF{(2^{2})}$ is shown in Figure \ref{fig:2-bit}. The left part of the figure shows a standard 2-bit integer multiplication with four output bits. To represent the result in $GF{(2^{m})}$, which can contain only two bits, the bits $r_{3}$ and $r_{2}$ are reduced in $GF(2^2)$. This result of such a reduction is shown on the right part of the figure. The input and output operands are represented using polynomials $A(x)$, $B(x)$ and $Z(x)$. The functions of $s_{0}$, $s_{1}$ and $s_{2}$ are  represented using polynomials in $GF(2)$: $s_{0}$=$a_{0}b_{0}$, $s_{1}$=$a_{1}b_{0}$+$a_{0}b_{1}$, and $s_{2}$=$a_{1}b_{1}$\footnote{For polynomials in $GF(2)$, "+" is computed modulo 2.}. Hence, $z_{0}$=$s_{0}$+$s_{2}$ and $z_{1}$=$s_{1}$+$s_{2}$. As a result, the coefficients of the multiplication are: $z_{0}$=$a_{0}b_{0}$+$a_{1}b_{1}$, $z_{1}$ = $a_{0}b_{1}$+$a_{1}b_{0}$+$a_{1}b_{1}$.  In digital circuits, partial products can be implemented using {\sc and} gates, and addition modulo 2 using {\sc xor} gates. Note that, unlike in the integer multiplication in $GF(2^m)$ circuits there is no carry out to the next bit. For this reason, as we can see in Figure \ref{fig:2-bit}, the function of each output bit is computed independently of other bits.

\subsection{Function Extraction}

\textit{Function extraction} is an arithmetic verification method proposed in \cite{ciesielski2015verification} for integer arithmetic circuits, in $\mathbb{Z}_{2^m}$. 
It extracts a unique bit-level polynomial function implemented by the circuit directly from its gate-level implementation. Extraction is done by \textit{backward rewriting}, i.e., transforming the polynomial representing encoding of the primary outputs (called the \textit{output signature}) into a polynomial at the primary inputs (the \textit{input signature}). This technique has been successfully applied to large integer arithmetic circuits, such as 512-bit integer multipliers. However, it cannot be directly applied to large $GF$ multipliers 
because of exponential size of the intermediate number of polynomial terms before cancellations during rewriting. 
Fortunately, arithmetic $GF(2^{m})$ circuits offer an inherent parallelism which can be exploited in backward rewriting.
In the rest of the paper, we show how to apply such parallel rewriting in $GF(2^{m})$ circuits while avoiding memory explosion experienced in integer arithmetic circuits.

\section{Preliminaries} \label{sec:preliminaries}

\subsection{Computer Algebraic model}

The circuit is modeled as a network of logic elements of arbitrary complexity including: basic logic gates (AND, OR, XOR, INV) and complex standard cell gates (AOI, OAI, etc.) obtained by synthesis and technology mapping. Instead of modeling Boolean operators using pseudo-Boolean equations, we use the algebraic models in $GF(2)$, i.e. modulo 2. For example, the pseudo-Boolean model of XOR($a,b$)=$a+b$ $- 2ab$ is reduced to $(a + b - 2ab)$ mod $2$ = $(a + b)$ mod $2$. The following algebraic equations are used to describe basic logic gates in $GF(2^{m})$, according to \cite{kalla:tcad13}:

\vspace{-4mm}
\begin{equation}
     \begin{aligned}
      \text{~~} &\\
       & \neg a = 1 + a \\
       & a \wedge b = a\cdot b \\
       & a \vee b = a + b + a\cdot b \\
       & a \oplus b = a + b
     \end{aligned}
\label{eq:boolean-poly}
\end{equation}

\subsection{Outline of the Approach}

Similarly to the work of \cite{ciesielski2015verification}, the computed function of the circuits is specified by two polynomials. The \textit{output signature} of a $GF(2^{m})$ multiplier, $Sig_{out} = \sum _{i=0} ^{m-1} z_i x^i$, with $z_i \in GF(2)$. The \textit{input signature} of a $GF(2^{m})$ multiplier, $Sig_{in}$ = $\sum _{i=0} ^{m-1} \mathbb{P}_i x^i$, with coefficients $\mathbb{P}_i \in GF(2)$ being product terms, with addition operation performed modulo 2 (e.g. ($a_0 b_0 + a_1 b_1$) mod 2). For a $GF(2^{m})$ multiplier, if the irreducible polynomial $P(x)$ is provided, $Sig_{in}$ is known. Our goal is to transform the output signature, $Sig_{out}$, using polynomial representation of the internal logic elements, into the input signature $Sig_{in}$ in $GF(2^m)$. The the goal of the verification problem is then to check if $Sig_{in}$ = $Sig_{out}$, expressed in the primary inputs. 

\textbf{Theorem 1:} \textit{Given a combinational $GF(2^m)$ arithmetic circuit, composed of logic gates, described by algebraic expressions (Eq. 1), input signature $Sig_{in}$ computed by backward rewriting is unique and correctly represents the function implemented by the circuit in $GF(2^m)$.}

\textbf{Proof:} The proof of correctness relies on the fact that each transformation step (rewriting iteration) is correct. That is, each internal signal is represented by an algebraic expression, which always evaluates to a {\it correct value} in $GF(2^{m})$. This is guaranteed by the correctness of the algebraic model in Eq. (\ref{eq:boolean-poly}), which can be proved easily by inspection. For example, the algebraic expression of \textit{XOR(a,b)} in $\mathbb{Z}_{2^m}$ is $a+b-2ab$. When implemented in $GF(2^{m})$, the coefficients in the expression must be in $GF(2)$. Hence, \textit{XOR(a,b)} in $GF{2^m}$ is represented by $a+b$. The proof of uniqueness is done by induction on $i$, the step of transforming polynomial $F_i$ into $F_{i+1}$. A detailed induction proof for expressions in $\mathbb{Z}_{2^m}$ is provided in \cite{ciesielski2015verification}.

\hfill $\square$

\textbf{Theorem 2:} \textit{Let the number of logic elements (polynomials) in a $GF(2^{m})$ multiplier be $n$. At each iterations, backward rewriting process generates $n$ internal expressions, $F_{0}, F_{1}, ..., F_{n-1}$, such that every expression $F_{i}$ $\in$ $GF(2^{m})$}.

\textbf{Proof:} Assuming that $F_{0}$=$Sig_{out}$ and each $F_{i} \in GF(2^{m})$, we prove that $F_{i+1}$ $\in$ $GF(2^{m})$. Each variable in $F_{i}$ represents output of some logic gate. During the rewriting process, this variable is substituted by a corresponding polynomial in Eq. (1). According to Theorem 1, resulting polynomial $F_{i+1}$ correctly represents the function $F_{i+1}$ $\in$ $GF(2^{m})$.

\hfill $\square$

\begin{algorithm}
\scriptsize
\caption{Backward Rewriting in $GF(2^{m})$}\label{alg:commonlogic}
\textbf{Input: Gate-level netlist of $GF(2^{m})$ multiplier}\\ 
\textbf{Input: Output signature $Sig_{out}$, and (optionally) input signature, $Sig_{in}$} \\
\textbf{Output: GF function of the design, and answer whether $Sig_{out}$==$Sig_{in}$}
\begin{algorithmic}[1]
\State $\mathcal{P}$=\{$p_{0},p_{1},...,p_{n}$\}: polynomials representing gate-level netlist
\State $F_{0}$=$Sig_{out}$
\For{each polynomial $p_{i}$ $\in \mathcal{P}$} 
\For{output variable $v$ of $p_{i}$ in $F_{i}$}
\State replace every variable $v$ in $F_{i}$ by the expression of $p_{i}$
\State $F_{i}$ $\rightarrow$ $F_{i+1}$
\For{each element/monomial $M$ in $F_{i+1}$}
\If {the coefficient of $M$\%2==0 \\~~~~~~~~~~~~~\textbf{or} $M$ is constant, $M$\%2==0}
\State remove $M$ from $F_{i+1}$
\EndIf
\EndFor
\EndFor
\EndFor \\
\Return $F_{n}$ and $F_{n}=?Sig_{in}$
\end{algorithmic}
\end{algorithm}

Theorems 1 and 2, together with the algebraic model in Eq. (1), provide the basis for polynomial reduction in backward rewriting in this work. This is described by Algorithm 1. Our method takes the gate-level netlist of a $GF(2^{m})$ multiplier as input and first converts each logic gate into equations using Eq. (1). The output signature $Sig_{out}$ is required to initialize the backward rewriting. The rewriting process starts with $F_{0}=Sig_{out}$, and ends when all the variables in $F_{i}$ are primary inputs. This is done by rewriting the polynomials representing logic elements in the netlist in a topological order \cite{ciesielski2015verification}. Each iteration includes two steps: Step 1) substitute the variable of the gate output using the expression in the inputs of the gate (Eq.1), and name the new expression $F_{i+1}$ (lines 3 - 6); Step 2) simplify the new expression by removing all the monomials (including constants) that evaluate to 0 in $GF(2)$ (line 3 and lines 7 - 10). The algorithm outputs the function of the design in $GF(2^m)$ after $n$ iterations, where $n$ is the number of gates in the netlist. The final expression $F_{n}$ can be used for functional verification, by checking if it matches the expected input signature (if provided). 

\begin{figure}[t] 
\begin{center}
\includegraphics[scale=0.37]{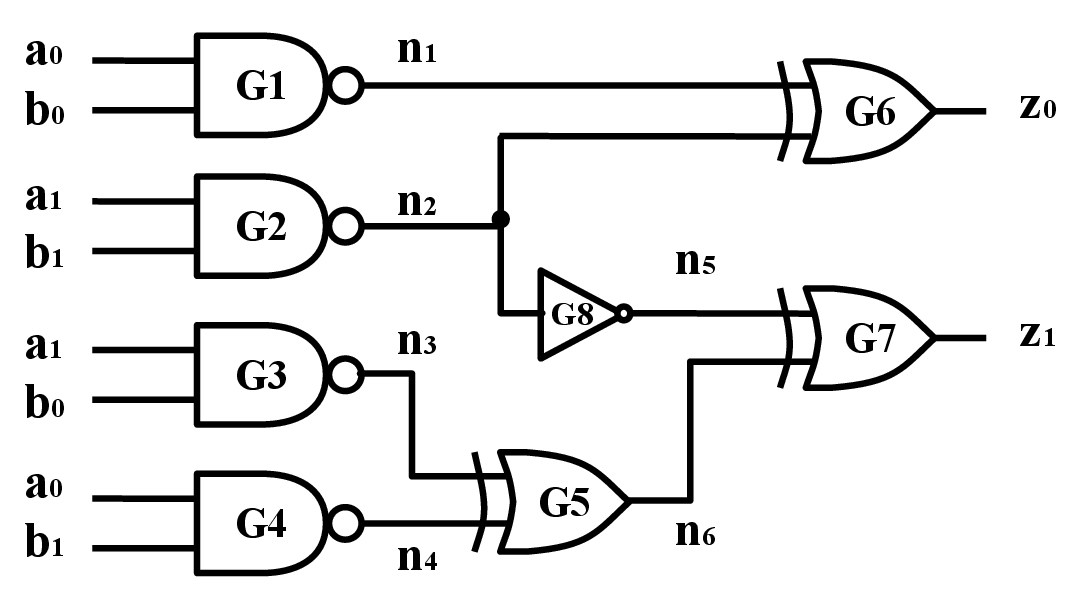}
\caption{The gate-level netlist of post-synthesized and mapped 2-bit multiplier over $GF(2^2)$. The irreducible polynomial $P(x)=x^{2}+x+1$.}
\vspace{-5mm}
\label{fig:netlist-2bit}
\end{center}
\end{figure}

\begin{figure}[]
\centering
\small
\begin{tabular}{|l|c|}
\hline
$Sig_{out}$: $F_0$=$z_0$+x$z_1$                                                &  Eliminating terms           \\ \hline
G7: $F_1$=$z_0$+x($n_5$+$n_6$)                                             & \textit{-}  \\ \hline
G6: $F_2$=$n_1$+$n_2$+x($n_5$+$n_6$)                                           & \textit{-}  \\ \hline
G5: $F_3$=$n_1$+$n_2$+x($n_3$+$n_4$+$n_5$)                                     & \textit{-}  \\ \hline
G8: $F_4$=$n_1$+$n_2$+x($n_3$+$n_4$+$n_2$+1)                                   & \textit{-}  \\ \hline
G4: $F_5$=$n_1$+$n_2$+x($n_2$+$n_3$+$a_0$$b_1$)+2x                          & \textit{2x} \\ \hline
G3: $F_6$=$n_1$+$n_2$+x($n_2$+$a_1$$b_0$+$a_0$$b_1$+1)                      & \textit{-}  \\ \hline
G2: $F_7$=$n_1$+$a_1$$b_1$+1+x($a_1$$b_1$+$a_1$$b_0$+$a_0$$b_1$)+2x      & \textit{2x} \\ \hline
G1: $F_8$=$a_0$$b_0$+$a_1$$b_1$+2+x($a_1$$b_1$+$a_1$$b_0$+$a_0$$b_1$) & \textit{2}  \\ \hline
$Sig_{in}$: $a_0$$b_0$+$a_1$$b_1$+x($a_1$$b_1$+$a_1$$b_0$+$a_0$$b_1$) & \textit{-}  \\ \hline
\end{tabular}
\caption{Function extraction of a 2-bit $GF$ multiplier shown in Figure 2 using backward rewiring from PO to PI.}
\vspace{-2mm}
\label{fig:rewriting}
\end{figure}

{\bf Example 1} (Figure \ref{fig:netlist-2bit}): We illustrate our method using a post-synthesized 2-bit multiplier in $GF(2^2)$, shown in Figure \ref{fig:netlist-2bit}. The irreducible polynomial is $P(x)$ = $x^{2}+x+1$. The output signature is $Sig_{out} = z_{0}$+$z_{1}x$, and input signature is $Sig_{in}=(a_{0}b_{0}$+$a_{1}b_{1}$)+($a_{1}b_{1}$+$a_{1}b_{0}$+$a_{0}b_{1}$)$x$. First, $F_{0}=Sig_{out}$ is transformed into $F_{1}$ using polynomial of gate $G7_7$, $z_{1}$=$n_{5}+n_{6}$. This expression is simplified to $F_{1}=z_{0}+n_{5}x+n_{6}x$. Then, the polynomials $F_{i+1}$ are successively derived from $F_{i}$ and checked for a possible reduction. The first reduction happens when $F_{4}$ is transformed into $F_{5}$, where $n_{4}$ (at gate $G_4$) is replaced by ($1+a_{0}b_{0}$). After simplification, a monomial $2x$ is identified and removed from $F_{5}$ since 2\%2=0. Similar reductions are applied during the transformations $F_{6} \rightarrow F_{7}$ and $F_{7} \rightarrow F_{8}$. Finally, the function of the design is extracted by Algorithm 1. A complete rewriting process is shown in Figure \ref{fig:rewriting}. We can see that $F_{8}=Sig_{in}$, which indicates that the circuit indeed implements the $GF(2^2)$ multiplication with $P(x)$=$x^{2}+x+1$.

An important observation is that the potential reductions take place only within the expression  associated with the same degree of polynomial ring ($Sig_{out}$ is a polynomial ring). In other words, the reductions happen independently in a logic cone of every output bit, independently of other bits, regardless of logic sharing between the cones. For example, the reductions in $F_{5}$ and $F_{7}$ are extracted from output $z_{1}$ only. Similarly, in $F_{8}$, the reduction is from $z_{0}$.

\textbf{Theorem 3:} \textit{Given a $GF(2^{m})$ multiplier with $Sig_{out}$ = $F_{0}$ = $z_{0}x^{0}$ + $z_{1}x^{1}$ + ... + $z_{m}x^{m}$; and $F_{i}$=$E_{0}x^{0}$ + $E_{1}x^{1}$ + ... + $E_{m}x^{m}$, where $E_{i}$ is an algebraic expression in $GF(2)$ obtained during rewriting. Then, the polynomial reduction is possible only within a single expression $E_{i}$, for $i$=1, 2, ..., m. }

\textbf{Proof:} Consider a polynomial $E_{i}x^{n_i}$+$E_{k}x^{n_k}$, where $E_{i}$ and $E_{k}$ are simplified in $GF(2)$. That is, $E_{i}=(e^1_{i} + e^2_{i} + ...$), and $E_{k}=(e^1_{k} + e^2_{k} + ...$). After simplifying each of the two polynomials, there are no common monomials between $E_{i}x^{n_i}$ and $E_{k}x^{n_k}$. This is because for any element, $e^l_{i}x^{n_i}$ $\neq$ $e^j_{k}x^{n_k}$, for any pairs of $(i, k)$ and $(l, j)$.

\hfill $\square$


\vspace{-3mm}
\section{Implementation} \label{sec:implementation}

\begin{figure}[!htb] 
\begin{center}
\includegraphics[scale=0.45]{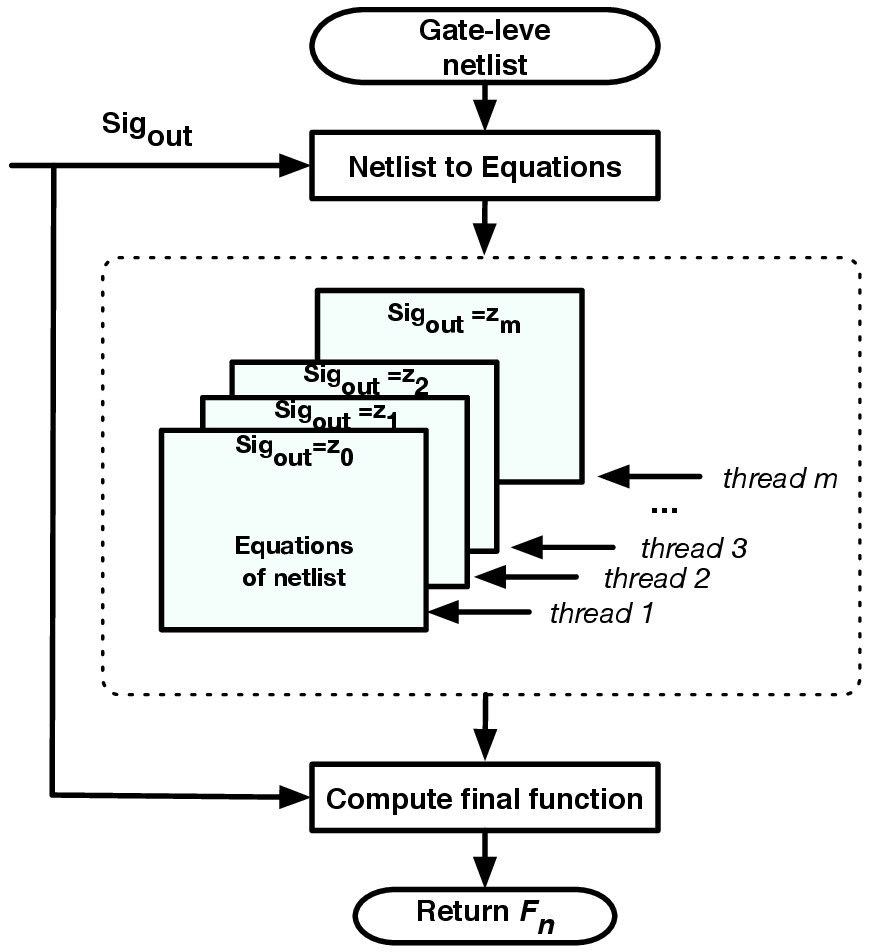}
\caption{Overview of parallel verification of GF multipliers.}
\vspace{-3mm}
\label{fig:flow}
\end{center}
\end{figure}

\begin{table*}[]
\scriptsize
\centering
\begin{tabular}{|c|c|c|c|c|c|c|c|c|}
\hline
\multicolumn{2}{|c|}{\textit{Mastrovito}} & \multicolumn{2}{c|}{\cite{kalla:dac2014}} & \multicolumn{5}{c|}{This work} \\ \hline
\multirow{2}{*}{Op size} & \multirow{2}{*}{\# equations} & \multirow{2}{*}{\begin{tabular}[c]{@{}c@{}}Runtime\\ (sec)\end{tabular}} & \multirow{2}{*}{\begin{tabular}[c]{@{}c@{}}Mem\\ (MB)\end{tabular}} & \multicolumn{4}{c|}{Runtime (sec)} & Mem* \\ \cline{5-9} 
 &  &  &  & \textit{T=5} & \textit{T=10} & \textit{T=20} & \textit{T=30} & \textit{T=1*} \\ \hline
32 & 5,482 & 0.83 & 3 & 1.90 & 1.54 & 0.95 & 1.09 & 10 MB \\ \hline
48 & 12,228 & 8.39 & 13 & 5.73 & 3.36 & 2.83 & 2.27 & 21 MB \\ \hline
64 & 21,814 & 28.90 & 21 & 11.08 & 7.88 & 6.87 & 6.74 & 37 MB \\ \hline
96 & 51,412 & 195.2 & 45 & 38.14 & 26.69 & 20.19 & 22.66 & 84 MB \\ \hline
128 & 93,996 & 924.3 & 91 & 91.67 & 62.68 & 54.99 & 56.76 & 152 MB \\ \hline
163 & 153,245 & 3546 & 161 & 192.6 & 137.5 & 120.7 & 113.1 & 248 MB \\ \hline
233 & 167,803 & 4933 & 168 & 294.1 & 212.7 & 180.1 & 170.6 & 270 MB \\ \hline
283 & 399,688 & 30358 & 380 & 890.7 & 606.5 & 549.7 & 529.8 & 642 MB \\ \hline
571 & 1628,170 & \textit{TO} & - & 7980 & 5038 &  \textit{MO} & \textit{MO} & 2.6 GB \\ \hline
\end{tabular}
\caption{Results of verifying Mastrovito multipliers using our parallel approach. $T$ is the number of threads. $TO$=Time out of 12 hours. $MO$=Memory out of 32 GB. \\(*\textit{T=1} shows the maximum memory usage of each thread.) }
\vspace{-3mm}
\label{tbl:mas}
\end{table*}
\begin{table*}[]
\scriptsize
\centering
\begin{tabular}{|c|c|c|c|c|c|c|c|c|}
\hline
\multicolumn{2}{|c|}{Montgomery} & \multicolumn{2}{c|}{\cite{kalla:dac2014}} & \multicolumn{5}{c|}{This work} \\ \hline
\multirow{2}{*}{Op size} & \multirow{2}{*}{\# equations} & \multirow{2}{*}{\begin{tabular}[c]{@{}c@{}}Runtime\\ (sec)\end{tabular}} & \multirow{2}{*}{\begin{tabular}[c]{@{}c@{}}Mem\\ (MB)\end{tabular}} & \multicolumn{4}{c|}{Runtime (sec)} & Mem* \\ \cline{5-9} 
 &  &  &  & \textit{T=5} & \textit{T=10} & \textit{T=20} & \textit{T=30} & \textit{T=1*} \\ \hline
32 & 4,352 & 1.98 & 3 & 3.49 & 2.16 & 1.31 & 2.08 & 8 MB \\ \hline
48 & 9,602 & 14.19 & 13 & 17.71 & 10.67 & 9.16 & 6.01 & 16 MB \\ \hline
64 & 16.898 & 63.48 & 21 & 44.86 & 30.57 & 28.3 & 27.22 & 27 MB \\ \hline
96 & 37,634 & 554.6 & 45 & 234.3 & 157.8 & 133.1 & 142.3 & 59 MB \\ \hline
128 & 66,562 & 1924 & 68 & 208.9 & 121.3 & 115.8 & 110.4 & 95 MB \\ \hline
163 & 107,582 & 12063 & 101 & 1615.7 & 1172.3 & 1094.9 & 1008.1 & 161 MB \\ \hline
233 & 219,022 & \textit{TO} & 168 & 722.3 & 564.8 & 457.7 & 479.8 & 301 MB \\ \hline
283 & 322,622 & \textit{TO} & 380 & 19745 & 17640 & 15300 & 14820 & 488 MB \\ \hline
\end{tabular}
\caption{Results of verifying \textit{Montgomery} multipliers using our parallel approach. $T$ is the number of threads. $TO$=Time out of 12 hours. $MO$=Memory out of 32 GB.\\(*\textit{T=1} shows the maximum memory usage of each thread.) }
\vspace{-3mm}
\label{tbl:mont}
\end{table*}

This section describes the implementation of our parallel verification method for Galois field multipliers. The overview of the proposed technique is shown in Figure \ref{fig:flow}. Our approach takes the gate-level netlist as input, and outputs the extracted function of the design. It includes four steps: 
\begin{enumerate}
\item Convert the gate-level netlist into algebraic equations. During this step, the gate-level netlist is translated into algebraic equations based on Eq.(1). The equations are levelized in topological order, to be rewritten by backward rewriting in the next step.
\item Split the output signature of $GF(2^{m})$ multipliers into $m$ polynomials with $Sig_{out\_i}$=$z_{i}$. These new \textit{signatures} are represented by $m$ equation files.
\item Split the function of $m$ output bits into $m$ separate functions, each to be processed by a separate thread using Algorithm 1. In contrast to work of \cite{ciesielski2015verification}, the internal expression of each output bit does not offer any polynomial reduction (\textit{monomial cancellations}) for other bits.
\item Compute the final function of the multiplier. Once the algebraic expression of each output bit in $GF(2)$ is computed, our method computes the final function by constructing the $Sig_{out}$ using the rewriting process in step 3.

\end{enumerate}
\begin{figure}[!htb]
\small
\centering
\begin{tabular}{|l|c|l|c|}
\hline
$Sig_{out0}$=$z_0$ & \begin{tabular}[c]{@{}c@{}}elim\end{tabular} & $Sig_{out1}$=x$\cdot$$z_1$ & \begin{tabular}[c]{@{}c@{}}elim\end{tabular} \\ \hline
G7: $z_0$ & \textit{-} & G7: x($n_5$+$n_6$) & - \\ \hline
G6: $n_1$+$n_2$ & \textit{-} & G6: x($n_5$+$n_6$) & - \\ \hline
G5: $n_1$+$n_2$ & \textit{-} & G5: x($n_3$+$n_4$+$n_5$) & - \\ \hline
G8: $n_1$+$n_2$ & \textit{-} & G8: x($n_3$+$n_4$+$n_2$)+x & - \\ \hline
G4: $n_1$+$n_2$ & \textit{-} & G4: x($n_2$+$n_3$+$a_0$$a_1$)+2x & 2x \\ \hline
G3: $n_1$+$n_2$ & \textit{-} & G3: x($n_2$+$a_1$$b_0$+$a_0$$b_1$)+x & - \\ \hline
G2: $n_1$+$a_1$$b_1$+1 & \textit{-} & G2: x($a_1$$b_1$+$a_1$$b_0$+$a_0$$b_1$)+2x & 2x \\ \hline
G1: $a_0$$b_0$+$a_1$$b_1$+2 & \textit{2} & G1: x($a_1$$b_1$+$a_1$$b_0$+$a_0$$b_1$) & - \\ \hline
\multicolumn{4}{|l|}{$Sig_{in}$=$a_0$$b_0$+$a_1$$b_1$+x($a_1$$b_1$+$a_1$$b_0$+$a_0$$b_1$)} \\ \hline
\end{tabular}
\caption{Parallel extraction of a 2-bit GF multiplier shown in Figure 2.}
\vspace{-2mm}
\label{fig:parallel}
\end{figure}

{\bf Example 2} (Figure \ref{fig:parallel}): We illustrate our parallel extraction method using the 2-bit multiplier in $GF(2^2)$ in Figure \ref{fig:netlist-2bit}. The output signature $Sig_{out}$ = $z_0$+$z_{1}x$ is split into two signatures, $Sig_{out0}=z_0$ and $Sig_{out1}=z_1$. Then, the rewriting process is applied to $Sig_{out0}$ and $Sig_{out1}$ in parallel. When $Sig_{out0}$ and $Sig_{out1}$ have been successfully extracted, the two signatures are merged as $Sig_{out0}$+$Sig_{out1}x$ resulting in the polynomial $Sig_{in}$. In Figure 3, we can see that elimination happens three times ($F_5$, $F_7$, and $F_8$). According to Theorem 3, we know that the elimination happens within each element in GF($2^n$). In Figure \ref{fig:parallel}, one elimination in $Sig_{out0}$ and two eliminations in $Sig_{out1}$ have been done independently, as shown earlier (Figure 3).

\section{Results}

The verification technique described in this paper was implemented in C++. It performs backward rewriting with variable substitution and polynomial reductions in Galois field, using the approach discussed in Sections \ref{sec:preliminaries} and \ref{sec:implementation}. 
The program was tested on a number of combinational gate-level $GF(2^{m})$ multipliers taken from \cite{kalla:tcad13}, including Montgomery multipliers \cite{koc1998montgomery} and Mastrovito multipliers \cite{sunar1999mastrovito}. The bit-width of the multipliers varies from 32 to 571 bits. The experiments of verifying Galois field multipliers using SAT, SMT, ABC \cite{abc-link} and Singular \cite{singular} have been presented in \cite{kalla:tcad13} and \cite{kalla:dac2014}. It shows that the rewriting technique performs significantly better than other techniques. Hence, in this work, we only compare our approach to those of \cite{kalla:tcad13} and \cite{kalla:dac2014}. Specifically, we compare our approach to the tool described in \cite{kalla:dac2014} on the same benchmark set. Our experiments were conducted on a PC with Intel(R) Xeon CPU E5-2420 v2 2.20 GHz x12 with 32 GB memory. As described in the next section, our technique is able to verify Galois field multipliers in multiple threads (up to 30 using our platform). In each thread, Algorithm 1 is applied on a single output bit. The number of threads is given as input to the tool. 

\subsection{Evaluation of Our Approach}

The experimental results of our approach and comparison against \cite{kalla:dac2014} are shown in Table \ref{tbl:mas} for gate-level Mastrovito multipliers with bit-width varying from 32 to 571 bits. These multipliers are directly mapped using ABC \cite{abc-link} without any optimization. The largest circuit includes more than 1.6 million gates. This is also the number of polynomial equations and the number of rewriting iterations (see Section 3). The results generated by the tool, presented in \cite{kalla:dac2014} are shown in columns 3 and 4. We performed four different series of experiments, with a different number of threads, $T$=5, 10, 20, and 30. The runtime results are shown in columns 6 to 8 and memory usage in column 9. The timeout limit (TO) was set to 12 hours and memory limit (MO) to 32 GB. The experimental results show that our approach provides on average 26.2x, 37.8x, 42.7x, and 44.3x speedup, for $T=$ 5, 10, 20, and 30 threads, respectively. Our approach can verify the multipliers up to 571 bit-wide multipliers in 1.5 hours, while that of \cite{kalla:dac2014} fails after 12 hours.

Note that the reported memory usage of our approach is the maximum memory usage {\it per thread}. This means that our tool experiences maximum memory usage with all $T$ threads running in the process; in this case, the memory usage is $T \cdot Mem$. This is why the 571-bit Mastrovito multipliers could be successfully verified with $T$ = 5 and 10, but failed with $T$ = 20 and 30 threads. For example, the peak memory usage of 571-bit Mastrovito multiplier with $T=20$ is $2.6 \times 20=52$ GB, which exceeds the available memory limit.

We also tested Montgomery multipliers with bit-width varying from 32 to 283 bits. These experiments are different than those in \cite{kalla:dac2014}. In our work, 
we first flatten the Montgomery multipliers before applying our verification technique. 
That is, we assume that only the positions of the primary inputs and outputs are known, without the knowledge of the internal structure or clear boundaries of the blocks inside the design. The results are shown in Table \ref{tbl:mont}. For 32- to 163-bit Montgomery multipliers, our approach provides on average a 9.2x, 15.9x, 16.6x, and 17.4x speedup, for $T=$ 5, 10, 20, and 30, respectively. Notice that \cite{kalla:dac2014} cannot verify the flattened Montgomery multipliers larger than 233 bits in 12 hours. 

In Table \ref{tbl:mont}, we observe that CPU runtime for verifying a 163-bit multiplier is greater than that of a 233-bit multiplier. This is because the computation complexity depends not only on the bit-width of the multipliers, but also on the irreducible polynomial $P(x)$. 

To analyze this dependency, we studied the effects of $P(x)$ on 4-bit multiplications implemented using different irreducible polynomials. The results are reported in Figure \ref{fig:study}). We can see that when $P(x)_{1}$=$x^{4}+x^{3}+1$, the longest logic paths for $z_{3}$ and $z_{0}$, include ten and seven products that need to be generated using XORs, respectively. However, when $P(x)_{2}$=$x^{4}+x+1$, the two longest paths, $z_{1}$ and $z_{2}$, have only seven and six products. This means that the GF($2^4$) multiplication requires \textbf{9} XOR operations using $P(x)_{1}$ and requires \textbf{6} XOR operations using $P(x)_{2}$. In other words, the gate-level implementation of the multiplier implemented using $P(x)_{1}$ has more gates compared to $P(x)_2$. In conclusion, we can see that irreducible polynomial $P(x)$ has significant impact on both design cost and the verification cost of the GF($2^m$) multipliers.

\begin{figure}[htb]
\centering
\small
\begin{tabular}{lllllll}
     &      &      & $a_3$   & $a_2$   & $a_1$   & $a_0$   \\
     &      &      & $b_3$   & $b_2$   & $b_1$   & $b_0$   \\ \hline
     &      &      & $a_{3}b_{0}$ & $a_{2}b_{0}$ & $a_{1}b_{0}$ & $a_{0}b_{0}$ \\
     &      & $a_{3}b_{1}$ & $a_{2}b_{1}$ & $a_{1}b_{1}$ & $a_{0}b_{1}$ &  \\
     & $a_{3}b_{2}$ & $a_{2}b_{2}$ & $a_{1}b_{2}$ & $a_{0}b_{2}$  &      &      \\
$a_{3}b_{3}$ & $a_{2}b_{3}$ & $a_{1}b_{3}$ & $a_{0}b_{3}$  &      &      &      \\ \hline
$s_6$   & $s_5$   & $s_4$   & $s_3$   & $s_2$   & $s_1$   & $s_0$  \\
\end{tabular}
\vspace*{5mm}
        \begin{minipage}{.5\linewidth}
      \centering
\begin{tabular}{cccc}
\multicolumn{4}{l}{$P(x)$=$x^{4}+x^{3}+1$}                                                                            \\
\multicolumn{1}{c|}{$s_3$} & \multicolumn{1}{c|}{$s_2$} & \multicolumn{1}{c|}{$s_1$} & $s_0$                     \\
\multicolumn{1}{c|}{$s_4$} & \multicolumn{1}{c|}{0}  & \multicolumn{1}{c|}{0} & $s_4$                     \\
\multicolumn{1}{c|}{$s_5$} & \multicolumn{1}{c|}{0}  & \multicolumn{1}{c|}{$s_5$}  & $s_5$                     \\
\multicolumn{1}{c|}{$s_6$} & \multicolumn{1}{c|}{$s_6$} & \multicolumn{1}{c|}{$s_6$}  & $s_6$                      \\ \hline
\multicolumn{1}{l|}{$z_3$} & \multicolumn{1}{l|}{$z_2$} & \multicolumn{1}{l|}{$z_1$} & \multicolumn{1}{l}{$z_0$}
\end{tabular}
    \end{minipage}%
    \begin{minipage}{.5\linewidth}
      \centering
\begin{tabular}{cccc}
\multicolumn{4}{l}{$P(x)$=$x^{4}+x+1$}                                                                            \\
\multicolumn{1}{c|}{$s_3$} & \multicolumn{1}{c|}{$s_2$} & \multicolumn{1}{c|}{$s_1$} & $s_0$                     \\
\multicolumn{1}{c|}{0} & \multicolumn{1}{c|}{0}  & \multicolumn{1}{c|}{$s_4$} & $s_4$                     \\
\multicolumn{1}{c|}{0} & \multicolumn{1}{c|}{$s_5$}  & \multicolumn{1}{c|}{$s_5$}  & 0                     \\
\multicolumn{1}{c|}{$s_6$} & \multicolumn{1}{c|}{$s_6$} & \multicolumn{1}{c|}{0}  & 0                      \\ \hline
\multicolumn{1}{l|}{$z_3$} & \multicolumn{1}{l|}{$z_2$} & \multicolumn{1}{l|}{$z_1$} & \multicolumn{1}{l}{$z_0$}
\end{tabular}    \end{minipage} 
\caption{Analysis of the computation complexity of Galois field multipliers with different irreducible polynomials using two 4-bit GF multiplications, which are implemented using $x^{4}+x^{3}+1$ and $x^{4}+x+1$.}
\vspace{-5mm}
\label{fig:study}
\end{figure}



\subsection{Runtime and Memory Tradeoff}

\begin{figure}[!hbt]
  \centering

    \includegraphics[width=0.32\textwidth]{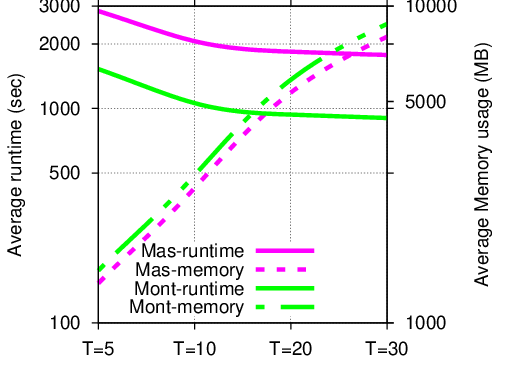}

\caption{Runtime and memory usage of our parallel verification approach as a function of number of threads $T$.}
\vspace{-5mm}
\label{fig:tradeoff}
\end{figure}

In this section, we discuss the tradeoff of runtime and memory usage of our approach. The plots in Figure \ref{fig:tradeoff} show how the average runtime and memory usage change with different number of threads. The vertical axis on the left is CPU runtime (in seconds), and on the right is memory usage (MB). Horizontal axis represents the number of threads $T$, ranging from 5 to 30. 
The runtime is significantly improved for $T$ between 5 and 15. However there is not much speedup when $T$ is greater than 20, most likely due to the memory management synchronization overhead between the threads.
Based on the experiments of Mastrovito multipliers (Table \ref{tbl:mas}), our approach is limited by the memory usage when the size of multiplier and $T$ are large. In our work, $T=20$ seems to be the best choice. Obviously, $T$ varies on different platform depending on the number of cores, and the memory.

\subsection{Verification of Synthesized GF Multipliers}

In \cite{cunxi:2016-tcad-verification}, the authors conclude that highly bit-optimized integer arithmetic circuits are harder to verify than their original, pre-synthesized netlists. This is because efficiency of the rewriting technique relies on the amount of cancellations between the different terms of the polynomial, and the cancellations strongly depend on the order in which signals are rewritten. A good ordering of signals is difficult to be achieved in highly bit-optimized circuits. 

In order to see the effect of synthesis on parallel verification of GF circuits, we applied our approach to {\it post-synthesized} Galois field multipliers with operands up to 409 bits (571-bit multipliers could not be synthesized in a reasonable time). 
We synthesized \textit{Mastrovito} and \textit{Montgomery} multipliers using $ABC$ tool \cite{abc-link}. We repeatedly used the commands \textit{resyn2} and \textit{dch}\footnote{"dch" is the most efficient bit-optimization function in ABC.} until $ABC$ could not reduce the number of levels or the number of nodes any more. The synthesized multipliers were mapped using a 14nm technology library. The verification experiments shown in Table \ref{tbl:synth} are performed by our tool with $T=20$ threads. Our tool was able to verify both 409-bit \textit{Mastrovito} and \textit{Montgomery} multipliers within just 13 minutes.
We observe that the Galois field multipliers are much easier to be verified after optimization. For example, the verification of a 283-bit Montgomery multiplier takes 15,300 seconds when $T=$20. After optimization, the runtime was just 169.2 seconds, which is 90x faster than verifying the original implementation. The memory usage is also reduced from 488 MB to 194 MB. In summary, in contrast to \cite{cunxi:2016-tcad-verification}, the bit-level optimization actually reduces the complexity of backward rewriting process. This is because extracting the function of an output bit of a GF multiplier depends only on the logic cone of this bit and does not require logic from other bits to be simplified (see Theorem 3). Hence, the complexity of function extraction is naturally reduced if logic cone is minimized. 

\begin{table}[!htb]
\scriptsize
\centering
\begin{tabular}{|c|c|c|c|c|}
\hline
\multirow{2}{*}{\textit{Op size}} & \multicolumn{2}{c|}{\textit{Mastrovito}} & \multicolumn{2}{c|}{\textit{Montgomery}} \\ \cline{2-5} 
 & Runtime & Mem & Runtime & Mem \\ \hline
64 & 4.25 s & 21 MB & 15.3 s & 38 MB\\ \hline
96 & 10.9 s & 44 MB & 40.5 s & 54 MB\\ \hline
128 & 28.9 s & 77 MB & 27.1 s & 78 MB\\ \hline
163 & 62.3 s & 123 MB & 205.2 s & 153 MB\\ \hline
233 & 134.8 s & 201 MB & 141.4 s & 199 MB\\ \hline
283 & 168.4 s & 198 MB & 169.2 s & 194 MB\\ \hline
409 & 775.6 s & 635 MB & 750.6 s & 597 MB\\ \hline
\end{tabular}
\caption{Runtime and memory usage of synthesized \textit{Mastrovito} and \textit{Montgomery} multipliers ($T$=20).}
\vspace{-5mm}
\label{tbl:synth}
\end{table}

\section{Conclusion}

In this paper, we present an algebraic functional verification technique of gate-level $GF(2^m)$ multipliers, in which verification is performed in bit-parallel fashion. The method is based on extracting a unique polynomial in Galois field of each output bit independently. We demonstrate that this method is able to verify an \textit{n}-bit GF multiplier in \textit{n} threads, while applying on pre- and post-synthesized \textit{Mastrovito} and \textit{Montgomery} multipliers up to 571 bits. The results show that our parallel approach gives average 44$\times$ and 17$\times$ speedup compared to the best existing algorithm. In addition, we analyze the effects of irreducible polynomial and synthesis on verification of GF($2^m$) multipliers.

\textbf{Acknowledgment:} This work was supported by an award from National Science Foundation, No. CCF-1319496 and No. CCF-1617708.

\scriptsize
\bibliographystyle{IEEEtran}
\bibliography{/Users/cunxiyu/Documents/TexShop/cunxi_bib/verification_ycunxi}
\end{document}